\documentclass{llncs}
\usepackage[T1]{fontenc}

\usepackage{graphicx}
\usepackage[inline]{enumitem}
\usepackage{adjustbox}
\newlength\myHeight 
\newlength\myWidth
\usepackage{booktabs}
\usepackage{adjustbox}
\usepackage{multirow}
\usepackage{svg}
\usepackage{amsmath}
\usepackage{orcidlink}
\usepackage{cleveref}
\usepackage{xstring}
\usepackage{float}
\usepackage{adjustbox}
\definecolor{mforestgreen}{HTML}{228B22}

\newcommand{\gptfouro}{GPT-4o}
\newcommand{\baseacceptedanswer}{accepted answer}

\newcommand{\acceptedanswers}{\StrSubstitute{\baseacceptedanswer}{answer}{answers}}

\newcommand{\expmine}{own}

\newcommand{\bert}{BERTScore}

\definecolor{myblue}{HTML}{0036c7}
\definecolor{mygreen}{HTML}{228B22}
\definecolor{myorange}{HTML}{FF5630}

\newcommand{\green}[1]{\textcolor{mygreen}{#1}}
\newcommand{\orange}[1]{\textcolor{myorange}{#1}}

\usepackage{ifthen}
\usepackage{ulem}
\newboolean{mocommentenabled}
\newboolean{onlysuggestedtext}
\newcommand{\enablemocomment}{\setboolean{mocommentenabled}{true}}

\newcommand{\mocomment}[2]{%
  \ifthenelse{\boolean{mocommentenabled}}{%
    \ifthenelse{\boolean{onlysuggestedtext}}{%
      \ifthenelse{\equal{#2}{}}{#1}{#2}%
    }{%
      \ifthenelse{\equal{#2}{}}{#1}{\sout{#1}\textcolor{blue}{#2}}%
    }%
  }{#1}%
}
\enablemocomment

\begin{document}
\title{LLM-Driven Personalized Answer Generation and Evaluation}
%
%
%
%
%
\author{
  Mohammadreza Molavi \inst{1} \orcidlink{0009-0006-0423-0729} \and
  Mohammadreza Tavakoli \inst{1} \orcidlink{0000-0002-7368-0794} \and
  Mohammad Moein \inst{1} \orcidlink{0000-0002-3285-8226} \and
  Abdolali Faraji \inst{1} \orcidlink{0000-0002-3557-9345} \and
  Gábor Kismihók \inst{1} \orcidlink{0000-0003-3758-5455}
}

\authorrunning{M. Molavi et al.}

\institute{Leibniz Information Centre for Science and Technology (TIB)
    \email{\{mohammadreza.molavi, reza.tavakoli, mohammad.moein, abdolali.faraji, gabor.kismihok\}@tib.eu}}

\maketitle              
\vspace{-1cm}
\begin{abstract}
  Online learning has experienced rapid growth due to its flexibility and accessibility. Personalization, adapted to the needs of individual learners, is crucial for enhancing the learning experience, particularly in online settings. A key aspect of personalization is providing learners with answers customized to their specific questions. This paper therefore explores the potential of \textit{Large Language Models} (\textit{LLMs}) to generate personalized answers to learners' questions, thereby enhancing engagement and reducing the workload on educators. To evaluate the effectiveness of LLMs in this context, we conducted a comprehensive study using the \textit{StackExchange} platform in two distinct areas: language learning and programming. We developed a framework and a dataset for validating automatically generated personalized answers. Subsequently, we generated personalized answers using different strategies, including 0-shot, 1-shot, and few-shot scenarios. The generated answers were evaluated using three methods: 1. \textit{\bert{}}, 2. LLM evaluation, and 3. human evaluation. Our findings indicated that providing LLMs with examples of desired answers (from the learner or similar learners) can significantly enhance the LLMs' ability to tailor responses to individual learners' needs.
  
\end{abstract}

\section{Introduction}
Online learning has transformed traditional education, offering flexibility and accessibility to learners worldwide \cite{gros2023future}. Its importance was underscored during the COVID-19 pandemic, which accelerated its adoption as a vital alternative to in-person learning \cite{Archambault_Leary_Rice_2022,Zhang_Carter_Qian_Yang_Rujimora_Wen_2022}. Personalization is an important element in online learning, improving efficiency and engagement \cite{liu2017data,gemeda2023inclusiveness}, and enabling learners to receive an education tailored to their individual needs \cite{tavakoli2020recommender,gros2023future}. A key aspect of personalization is providing tailored answers to learners’ questions, offering individualized guidance that enhances their learning \cite{kokku2018augmenting,pratama2023revolutionizing}.

Accordingly, efforts have been made to provide personalized answers to learners' questions \cite{yang2024ta,zhang2024design}. Yang et al. \cite{yang2024ta} developed \textit{YA-TA}, a virtual teaching assistant aimed at providing personalized support in large classes. It used a framework that combines instructor and learner knowledge to generate relevant responses. Classroom experiments showed the framework's effectiveness in delivering contextually accurate support. Moreover, Zhang \cite{zhang2024design} proposed an intelligent Q\&A system for online education platforms using \textit{Natural Language Processing} (\textit{NLP}). The system employed techniques like named entity recognition, sentiment analysis, and semantic understanding to effectively analyze and respond to user queries.

The emergence of \textit{Large Language Models} (\textit{LLMs}) has stimulated research into automatic question answering \cite{tan2023can,hasan2023exploratory,siddiq2024quality}. For instance, Tan et al. \cite{tan2023can} evaluated LLMs on 190,000 complex Knowledge-Based (KB) questions using the \textit{CheckList} framework, identifying limitations in handling real-world KB queries. As another example, Hasan et al. \cite{hasan2023exploratory} analyzed GPT-3's impact on online learning communities like \textit{StackOverflow}, examining linguistic strengths, problem-solving accuracy, and effects on user participation. Despite these efforts, the development of methods specifically customized to personalized answering remains an underexplored area. These methods not only enhance the learning experience by providing personalization \cite{pratama2023revolutionizing} but also significantly reduce the time burden on educators tasked with answering learners' questions \cite{kokku2018augmenting}.

\begin{figure}[t]
  \centering
  \adjustbox{max width=\textwidth,max height=0.45\textheight, center,keepaspectratio}{\includegraphics[width=\textwidth]{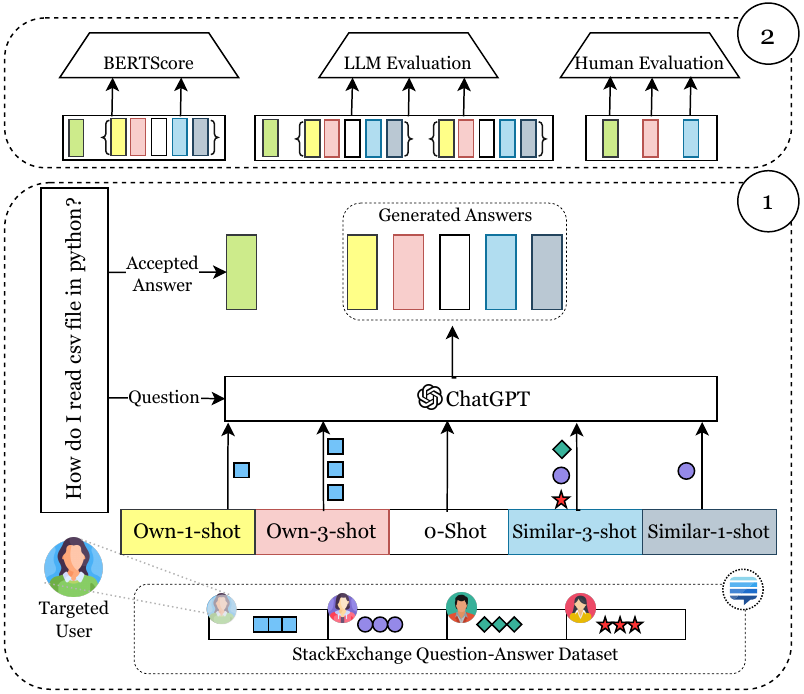}}
  \caption{The proposed pipeline for personalized answer generation consists of two phases: \textbf{(1)} scenario creation based on contextual answers. \textbf{(2)} evaluation using \bert{}, LLM evaluation, and human evaluation.}
  \label{fig:toc}
\end{figure}

Therefore, this paper aims to (1) develop a framework for evaluating personalized answer generation and (2) investigate the performance of different strategies, such as 0-shot, 1-shot, and few-shot scenarios, in generating personalized answers.
\Cref{fig:toc} depicts our pipeline, which includes the following steps:
\begin{enumerate}
  \item \textbf{\textit{Collecting a dataset.}} Our study required a dataset where learners explicitly expressed their satisfaction with answers. To address this, we utilized the StackExchange platform, where users pose questions and indicate whether an answer was accepted (desired) by them or not.
  \item \textbf{\textit{Generating various answers.}} We developed different methods to investigate how personalization could improve the baseline answers (0-shot scenario) which were generated by an LLM.
  \item \textbf{\textit{Evaluating the generated answers.}} To assess the quality of generated answers, we employed a three-phase evaluation approach: 1) calculating \bert{} similarity to desired answers, 2) using an LLM to evaluate the generated answers, and 3) obtaining human expert blind decisions on the similarity of the generated answers to the desired ones.
\end{enumerate}

\section{Method}
This study focuses on two key educational domains: English language learning and programming \cite{messer2024automated,boguslawski2024programming,shadiev2024review,choi2024english}. For these domains, we collected the necessary dataset from the StackExchange platform and developed strategies to generate various personalized responses.

\textbf{Dataset.} For creating our dataset, we relied on \textit{StackExchange}\footnote{\url{https://data.stackexchange.com/}}. While StackExchange is not a traditional learning platform, it is a recognized educational technology platform and plays a significant role in lifelong, informal, workplace, and problem-based learning scenarios \cite{stack,prosus}. We used a publicly available dataset on \textit{BigQuery}\footnote{\url{https://bit.ly/stackdump}}, which included a StackOverflow dump up to the end of 2022. To ensure diverse programming topics (e.g., backend, frontend, data science), we selected all questions from 2022 tagged with Python or JavaScript. We focused on users who had asked at least four questions to enable 0-shot, 1-shot, and few-shot evaluations. For these users, we collected all questions with both an accepted answer and at least one other response. The accepted answers served as benchmarks for the users' preferred outcomes. This strategy enabled us to evaluate how effectively our generated answers aligned with learners' needs, as users understood the accepted answer, knew how to apply it, and preferred it over other alternatives. For the English language, due to the limited availability of questions, we applied the same strategy to collect questions posted from 2018 onward using the StackExchange Data Explorer tool. The dataset specifications are as follows: Python programming included \textit{553} users and \textit{3,379} questions, JavaScript programming had \textit{276} users and \textit{1,630} questions, and English learning comprised \textit{341} users and \textit{1,564} questions.

\textbf{Personalized Answer Generation.} We explored personalized answer generation using well-established prompting scenarios by providing \gptfouro{} with zero, one, or three sample desired answers as input, which are detailed below:

\noindent\begin{enumerate}[label=(\arabic*)]
   \item \textbf{\textit{0-shot.}} We did not provide any sample answers in this scenario, establishing a baseline for comparison with scenarios that include sample answers for personalization.
\item \textbf{\textit{Own-1-shot.}} We enriched the prompt with the accepted answer from the user's previous question.
  \item \textbf{\textit{Own-3-shot.}} Similar to the previous scenario, we incorporated the accepted answers from the user's three previous questions into the prompt.
  \item \textbf{\textit{Similar-1-shot.}} To mimic real-world scenarios, where many users have not previously asked questions, we opted to utilize answers from similar users. In this context, we provided a randomly selected accepted answer to a question in the same area in this scenario.
  \item \textbf{\textit{Similar-3-shot.}} Accordingly, we incorporated three randomly selected answers from similar users into this scenario.
\end{enumerate}

\begin{figure}[t]
  \centering
  \begin{adjustbox}{max width=.9\textwidth, keepaspectratio}
  \includegraphics{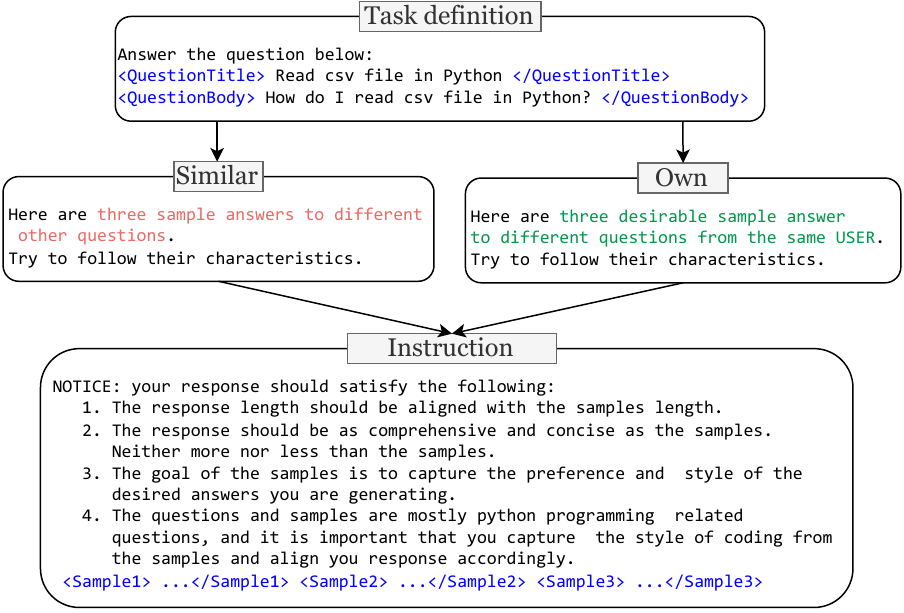}
  \end{adjustbox}
  \caption{Our 3-shot prompt scenarios for Python questions, which used sample accepted answers from \orange{similar users} (\orange{similar} scenario) and \green{the same users} (\green{\expmine{}} scenario).}
  \label{fig:prompt}
\end{figure}

Finally, we queried \gptfouro{} with prompts consisting of three parts: task, context, and instructions (\Cref{fig:prompt}). This prompt design follows best practices, including \textit{explicit task definition}, \textit{structured input and output}, and \textit{contextual guidance} \cite{Sahoo_Singh_Saha_Jain_Mondal_Chadha_2024}. Notably, the \textit{0-shot} scenario only used the task part.

\section{Evaluation}
Answers generated by \gptfouro{} were evaluated using \bert{} \cite{Zhang_Kishore_Wu_Weinberger_Artzi_2020}, LLM evaluation, and human evaluation to assess the quality and personalization alignment with desired \acceptedanswers{}.

\textbf{\bert{}.} This metric \cite{Zhang_Kishore_Wu_Weinberger_Artzi_2020} is widely used to evaluate text quality using contextual embeddings, making it robust to syntactic differences. It correlates well with human judgment, making it a common choice for assessing text generation models. You can find the \bert{} calculation results in \Cref{tab:performance_metrics}. It can be seen that the 3-shot scenario using learners' previous desired answers achieved the highest scores across all areas, suggesting improved alignment with user needs. All scenarios outperformed the 0-shot approach, indicating that providing context---such as sample desired answers from the same or similar users---enhances response generation.

\begin{table}[t]
  \Large
  \centering
  \begin{adjustbox}{max width=\textwidth}
    \begin{tabular}{l|c|c|c|c|c}
      \toprule
                          & \textbf{0-shot} & \textbf{Similar-1-shot} & \textbf{Similar-3-shot} & \textbf{Own-1-shot} & \textbf{Own-3-shot} \\
      \midrule
      \textbf{Python}     & 64.99\%            & 68.08\%                   & 68.31\%                     & 68.20\%               & \textbf{68.50}\%        \\
      \textbf{JavaScript} & 67.34\%            & 69.56\%                   & 69.80\%                     & 69.89\%               & \textbf{70.20}\%        \\
      \textbf{English}    & 52.04\%            & 52.14\%                   & 52.32\%                     & 52.31\%               & \textbf{52.58}\%        \\
      \bottomrule
    \end{tabular}
  \end{adjustbox}
  \caption{\bert{} evaluation results.}
  
  \label{tab:performance_metrics}
\end{table}

\textbf{LLM evaluation.} With the help of \gptfouro{}, this method aimed to assess the generated answers by identifying which one most closely resembled and aligned with the user's accepted answer. Therefore, \gptfouro{} was provided with the (1) original question and the user’s accepted (desired) answer, and (2) two generated answers from the five defined scenarios. This resulted in ten pairwise evaluations per question. To mitigate order bias \cite{shi2024judging}, we randomized the ordering of the generated answers in the prompt. To select among the answers, the following criteria were provided to \gptfouro{}:

\begin{enumerate}[label=(\arabic*)]
  \item \textbf{Coverage}: Ensure that all question requirements are met.
  \item \textbf{Detail consistency}: Match the detail level of the accepted answer.
  \item \textbf{Style consistency}: Match the style of the question and accepted answer.
  \item \textbf{Coding consistency} (for programming): Match methods with the accepted answer.
  \item \textbf{Correctness}: Ensure correctness of the generated answer.
\end{enumerate}

For each pair of scenarios, we analyzed GPT's selections across all questions. \Cref{tab:comparison_metrics} shows the most frequently chosen scenario along with its selection frequency for different scenario pairs. Consistent with the \bert{} findings, the own-3-shot and own-1-shot scenarios achieved superior results, reinforcing the notion that providing sample desired answers from the same users can enhance the likelihood of generating preferred responses. Moreover, \Cref{tab:comparison_metrics} clearly distinguishes the 0-shot scenario from the others, demonstrating a significant improvement when desired answers from the same or similar users are provided to \gptfouro{}.

\begin{table}[h]
  \centering
  \begin{adjustbox}{width=\textwidth, height=0.1\textheight}
    \begin{tabular}{l@{\hskip 2mm} l@{\hskip 2mm} l l l l l}
      \toprule
      \multicolumn{3}{c}{\textbf{Scenario pairs}} & \multicolumn{3}{c}{GPT Selected Scenario (Frequency in \%)}  \\
        \cmidrule(lr){4-6}
       &  &  & \textbf{Python} & \textbf{JavaScript} & \textbf{English}\\
      \midrule
      \textbf{own-3-shot} & vs. & \textbf{similar-3-shot} & own-3-shot (51.5)    & own-3-shot (50.38)      & own-3-shot (53)      \\
      \textbf{own-3-shot} & vs. & \textbf{similar-1-shot}  & own-3-shot (53.86)   & own-3-shot (56.81)      & own-3-shot (57.4)    \\
      \textbf{own-3-shot} & vs. & \textbf{own-1-shot}      & own-3-shot (53.45)   & own-3-shot (56.23)      & own-1-shot (50.61)   \\
      \textbf{own-3-shot} & vs. & \textbf{0-shot}          & own-3-shot (80.08)   & own-3-shot (83.77)      & own-3-shot (72.24)   \\
      \textbf{own-1-shot} & vs. & \textbf{similar-1-shot}  & own-1-shot (51.29)     & own-1-shot (50.58)        & own-1-shot (54.77)     \\
      \textbf{own-1-shot} & vs. & \textbf{similar-3-shot}  & similar-3-shot (50.16)     & similar-3-shot (56.63)        & own-1-shot (50.23)     \\
      \textbf{own-1-shot} & vs. & \textbf{0-shot}          & own-1-shot (79.05)     & own-1-shot (79.65)        & own-1-shot (70.22)     \\
      \textbf{similar-1-shot} & vs. & \textbf{0-shot}        & similar-1-shot (79.79) & similar-1-shot (81)       & similar-1-shot (70.12) \\
      \textbf{similar-1-shot} & vs. & \textbf{similar-3-shot}  & similar-3-shot (53.2)  & similar-3-shot (56.34)    & similar-3-shot (55.47) \\
      \textbf{similar-3-shot} & vs. & \textbf{0-shot}      & similar-3-shot (79.59) & similar-3-shot (84.33) & similar-3-shot (72.62) \\
      \bottomrule
    \end{tabular}
  \end{adjustbox}
  \caption{LLM evaluation results.}
  \label{tab:comparison_metrics}
\end{table}

\textbf{Human Evaluation. } In the final evaluation step, we involved at least two experts per domain to choose between pairs of generated answers based on their similarity to the accepted answer. We used the \textit{own-3-shot} and \textit{similar-3-shot} scenarios, identified as optimal in prior evaluations. This process aimed to gather expert opinions, as human judgment is the standard for qualitative assessments. To ensure fair labeling, we developed an application that allows experts to blindly vote on which generated answer best matches the user’s desired answer, with an option to skip if unsure. Due to the time-intensive nature of this process, taking approximately ten minutes per question, we gathered 100 labeled questions from each of the domains. This evaluation also served to verify the validity of results made by both \bert{} and LLM evaluations. Experts preferred the own-3-shot scenario in 67\% of Python, 68\% of JavaScript, and 70\% of English cases, aligning with earlier findings. We also asked the experts to justify their choice of answers in the form of open-ended text. The most frequently mentioned reasons were: (1) citing references from dictionaries and/or the history of terms in the English language domain, (2) employing a similar coding style (e.g., using comments and functions) and applying the same methods within the programming domain, and (3) offering a comparable level of detail (e.g., providing step-by-step explanations).

\section{Conclusion}
The emphasis on personalization in online learning has led to solutions that tailor the experience to individual needs, including generating answers specific to each learner. By leveraging LLMs, educators can save time while offering more personalized support. This study highlights the potential of LLMs to generate personalized answers in online learning environments.

By developing a framework and a dataset from StackExchange for evaluating personalized answers, we provided opportunities for future studies in this area. Our investigation of different strategies for generating personalized answers showed that providing \gptfouro{} with sample desired answers from the same learner or similar learners would increase the chance of generating more preferred answers. Our results were based on leveraging various evaluation strategies, including \bert{}, LLM evaluation, and human evaluation.

Our work-in-progress study showed promising results as an initial step, demonstrating that even without complex processing, LLMs are able to capture certain aspects of personalization from previous users' answers. However, further research is needed to explore the limitations and challenges of using LLMs for personalized question answering. Future work should focus on improving answer quality, addressing biases in LLM training data, and investigating the long-term impact on learning outcomes. Additionally, exploring more datasets and domains is necessary to validate the solution's applicability across various contexts.

%
%
\bibliographystyle{splncs04}
\bibliography{References}

\begin{thebibliography}{10}
\providecommand{\url}[1]{\texttt{#1}}
\providecommand{\urlprefix}{URL }
\providecommand{\doi}[1]{https://doi.org/#1}

\bibitem{Archambault_Leary_Rice_2022}
Archambault, L., Leary, H., Rice, K.: Pillars of online pedagogy: A framework for teaching in online learning environments. Educational Psychologist  \textbf{57}(3),  178–191 (Jul 2022). \doi{10.1080/00461520.2022.2051513}

\bibitem{boguslawski2024programming}
Boguslawski, S., Deer, R., Dawson, M.G.: Programming education and learner motivation in the age of generative ai: student and educator perspectives. Information and Learning Sciences  (2024)

\bibitem{choi2024english}
Choi, L.J.: English as an important but unfair resource: University students’ perception of english and english language education in south korea. Teaching in Higher Education  \textbf{29}(1),  144--158 (2024)

\bibitem{gemeda2023inclusiveness}
Gemeda, M.G.C., Yadavalli, P.K., Legesse, M.Y.K., Yadessa, M.A.A., Hirpa, M.D.A., Luta, M.R.B.: INCLUSIVENESS IN EDUCATION. Laxmi Book Publication (2023)

\bibitem{gros2023future}
Gros, B., Garc{\'\i}a-Pe{\~n}alvo, F.J.: Future trends in the design strategies and technological affordances of e-learning. In: Learning, design, and technology: An international compendium of theory, research, practice, and policy, pp. 345--367. Springer (2023)

\bibitem{hasan2023exploratory}
Hasan, M.M., Hasan, M., Nayeem, J.U.: An exploratory analysis of community-based question-answering platforms and gpt-3-driven generative ai: Is it the end of online community-based learning?  (2023)

\bibitem{kokku2018augmenting}
Kokku, R., Sundararajan, S., Dey, P., Sindhgatta, R., Nitta, S., Sengupta, B.: Augmenting classrooms with ai for personalized education. In: 2018 IEEE international conference on acoustics, speech and signal processing (ICASSP). pp. 6976--6980. IEEE (2018)

\bibitem{liu2017data}
Liu, D.Y.T., Bartimote-Aufflick, K., Pardo, A., Bridgeman, A.J.: Data-driven personalization of student learning support in higher education. Learning analytics: Fundaments, applications, and trends: A view of the current state of the art to enhance e-learning pp. 143--169 (2017)

\bibitem{messer2024automated}
Messer, M., Brown, N.C., K{\"o}lling, M., Shi, M.: Automated grading and feedback tools for programming education: A systematic review. ACM Transactions on Computing Education  \textbf{24}(1),  1--43 (2024)

\bibitem{pratama2023revolutionizing}
Pratama, M.P., Sampelolo, R., Lura, H.: Revolutionizing education: harnessing the power of artificial intelligence for personalized learning. Klasikal: Journal of education, language teaching and science  \textbf{5}(2),  350--357 (2023)

\bibitem{prosus}
Prosus: {Stack Exchange}. \url{https://en.wikipedia.org/wiki/Prosus} (2025)

\bibitem{Sahoo_Singh_Saha_Jain_Mondal_Chadha_2024}
Sahoo, P., Singh, A.K., Saha, S., Jain, V., Mondal, S., Chadha, A.: A systematic survey of prompt engineering in large language models: Techniques and applications (arXiv:2402.07927) (Feb 2024), \url{http://arxiv.org/abs/2402.07927}, arXiv:2402.07927 [cs]

\bibitem{shadiev2024review}
Shadiev, R., Yu, J.: Review of research on computer-assisted language learning with a focus on intercultural education. Computer Assisted Language Learning  \textbf{37}(4),  841--871 (2024)

\bibitem{shi2024judging}
Shi, L., Ma, W., Vosoughi, S.: Judging the judges: A systematic investigation of position bias in pairwise comparative assessments by llms. arXiv preprint arXiv:2406.07791  (2024)

\bibitem{siddiq2024quality}
Siddiq, M.L., Roney, L., Zhang, J., Santos, J.C.D.S.: Quality assessment of chatgpt generated code and their use by developers. In: Proceedings of the 21st International Conference on Mining Software Repositories. pp. 152--156 (2024)

\bibitem{tan2023can}
Tan, Y., Min, D., Li, Y., Li, W., Hu, N., Chen, Y., Qi, G.: Can chatgpt replace traditional kbqa models? an in-depth analysis of the question answering performance of the gpt llm family. In: International Semantic Web Conference. Springer (2023)

\bibitem{tavakoli2020recommender}
Tavakoli, M., Hakimov, S., Ewerth, R., Kismih{\'o}k, G.: A recommender system for open educational videos based on skill requirements. In: 2020 IEEE 20th International Conference on Advanced Learning Technologies (ICALT). pp.~1--5. IEEE (2020)

\bibitem{stack}
Wikipedia: {Stack Exchange}. \url{https://en.wikipedia.org/wiki/Stack\_Exchange} (2025)

\bibitem{yang2024ta}
Yang, D., Lee, S., Kim, M., Won, J., Kim, N., Lee, D., Yeo, J.: Ya-ta: Towards personalized question-answering teaching assistants using instructor-student dual retrieval-augmented knowledge fusion. arXiv preprint arXiv:2409.00355  (2024)

\bibitem{Zhang_Carter_Qian_Yang_Rujimora_Wen_2022}
Zhang, L., Carter, R.A., Qian, X., Yang, S., Rujimora, J., Wen, S.: Academia’s responses to crisis: A bibliometric analysis of literature on online learning in higher education during covid‐19. British Journal of Educational Technology  \textbf{53}(3),  620–646 (May 2022). \doi{10.1111/bjet.13191}

\bibitem{Zhang_Kishore_Wu_Weinberger_Artzi_2020}
Zhang, T., Kishore, V., Wu, F., Weinberger, K.Q., Artzi, Y.: Bertscore: Evaluating text generation with bert (arXiv:1904.09675) (Feb 2020). \doi{10.48550/arXiv.1904.09675}, \url{http://arxiv.org/abs/1904.09675}, arXiv:1904.09675 [cs]

\bibitem{zhang2024design}
Zhang, Y.: Design of an intelligent q\&a system for online education platform based on natural language processing technology. Journal of Electrical Systems  \textbf{20}(6s),  2135--2145 (2024)

\end{thebibliography}
\end{document}